\documentstyle[12pt,epsfig]{article} 
\begin{document}
\title{Gravity--Induced Interference and Continuous Quantum Measurements.}
\author{ A. Camacho
\thanks{email: acamacho@aip.de} \\
Astrophysikalisches Institut Potsdam. \\
An der Sternwarte 16, D--14482 Potsdam, Germany.}

\date{}
\maketitle

\begin{abstract}
Gravity--induced quantum interference is a remarkable effect that has already 
been confirmed experimentally, and it is a phenomenon in which quantum mechanics 
and gravity play simultaneously an important role. 
Additionally, a generalized version of this interference experiment could offer the possibility to confront against measurement outputs 
one of the formalisms that claim to give an explanation to the so called quantum measurement problem, 
namely the restricted path integral formalism. In this work we will analyze a possible extension of Colella, Overhauser, and Werner experiment 
and find that in the context of the res\-tric\-ted path integral formalism we obtain 
new in\-ter\-fe\-rence terms that could be measured in an extended version of this experimental 
cons\-truc\-tion. These new terms not only show, as in the first experiment, that at the quantum level gravity 
is not a purely geometric effect, it still depends on mass, but also show that interference 
does depend on some parameters that appear in the restricted path integral formalism, thus 
offering the possibility to have a testing framework for its theoretical predictions. 

\end{abstract}

\newpage
\section{Introduction.}
\bigskip

In the context of classical mechanics mass $m$ does not appear in the motion equation of a particle trajectory, 
gravity in classical mechanics could be then considered a purely geometric effect. 
But this situation does not happen in quantum theory, here mass does no longer cancel, 
instead it always appears in the combination $m/\hbar$ [1]. 

In order to detect nontrivial quantum--mechanical effects of gravity the co\-rres\-ponding 
experimental device must be able to analyze effects in which $\hbar$ appears explicitly.
The free fall of an elementary particle does not allow the study of this issue, because 
Ehrenfest theorem (where $\hbar$ does not appear) suffices to account for it. 
Even the red shift effect of a photon in a gravitational field [2] does not offer this possibility, here 
only a frequency shift is measured, and once again $\hbar$ does not appear explicitly. 

But in 1975 a neutron interferometer was used in order to detect the quantum--mechanical 
phase shift of neutrons caused by their interaction with Earth's gravitational field [3]. 
A nearly monoenergetic beam of thermal neutrons is split in two parts, each one of these new 
beams follow different paths in Earth's gravitational field, and then they are brought together. 
A gravity--induced phase shift emerges which depends on $(m/\hbar)^2$, being $m$ the neutron mass. 
It is interesting to comment that this phase shift due to  gravity is seen to be verified to well within one percent. 
An important point in the theoretical analysis of this phase shift lies in the 
fact that the wave function of the thermal neutrons is much smaller than the macroscopic dimension 
of the loop formed by the two alternate paths, therefore the concept of classical trayectory can 
be used in the derivation of the expression for the phase shift [1]. 
At this point we may wonder how this expression has to be reformulated if we 
do not consider this last condition.

It would be interesting to relate this gravity--induced interference with one of the most important problems in Quantum Theory (QT), namely the so called quantum 
measurement problem. In the attempts to solve this very old conundrum we may find several approaches, 
one of them is the so called Restricted Path Integral Formalism (RPIF) [4]. This formalism 
explains a continuous quantum measurement with the introduction of a restriction on 
the integration domain of the corresponding path integral. This last condition can also be reformulated in terms 
of a weight functional that has to be considered in the path integral. 

Let us explain this point a little bit better, and suppose that we have a particle which shows one--dimensional movement. 
The amplitude $A(q'', q')$ for this particle to move from the point $q'$ to the point $q''$ is called propagator. 
It is given by Feynman [5] 

\begin{equation}
A(q'', q') = \int d[q]exp({i\over \hbar}S[q]),
\end{equation}

\noindent here we must integrate over all the possible trajectories $q(t)$ and $S[q]$ is the action of the system, which is 
defined as

\begin{equation}
S[q] = \int_{t'}^{t''}dtL(q, \dot{q}).
\end{equation}

Let us now suppose that we perform a continuous measurement of the position of this particle, 
such that we obtain as result of this measurement process a certain output $a(t)$. In other words, the measurement process gives the value $a(t)$ 
for the \-coor\-di\-na\-te $q(t)$ at each time $t$, and this output has associated a certain error $\Delta a$, which is determined by the 
experimental resolution of the measuring device. The amplitude $A_{[a]}(q'', q')$ can be now thought of as a probability amplitude for the continuous measurement process to give the result $a(t)$. 
Taking the square modulus of this amplitude allows us to find the probability density for different measurement outputs.

Clearly, the integration in the Feynman path--integral should be restricted to those trajectories that match with the experimental output. 
RPIF says that this 
condition can be introduced by means of a weight functional $\omega_a[q]$ [4]. 
This means that expression (1) 
becomes now under a continuous measurement process

\begin{equation}
A_a = \int d[q]\omega_a[q]exp(iS[q]).
\end{equation}

The more probable the trajectory $[q]$ is, according to the output $a$, the bigger that $\omega_a[q]$ becomes [4]. This means that the value of $\omega_a[q]$ is approximately one for all trajectories $[q]$ that agree with the measurement 
output $a$ and it is almost 0 for those that do not match with the result of the experiment. 
Clearly, the weight functional contains 
all the information about the interaction between measuring device and measured system. 

This formalism has been employed in several situations, i.e., the analysis of 
the response of a gravitational wave antenna of Weber type [4], 
the measuring process of a gravitational wave in a laser--interferometer [6], 
or even to explain the emergence of the classical concept of time [7]. 

But even though there are already some theoretical predictions that could render a 
framework which could allow us to confront RPIF against experimental outputs [8], 
it is also true that more results are needed in this direction. 

The idea in this work is to derive some results using RPIF 
in the context of a generalized gravity--induced interference scenario. 
We will consider a possible ge\-ne\-ra\-li\-zation of Colella, Overhauser, and Werner (COW) experiment, namely the condition around the size of the wave 
packets (they must be smaller than the dimensions of the loop of the two alternate paths) will not be introduced anymore, i.e., we can 
not analyze this in\-ter\-fe\-rence process using the classical trayectories of the split beams.
Each one of these split beams will also be subject to the continuous monitoring of its vertical coordinate, 
additionally we introduce the possibility that the whole experimental device 
could be mounted on an accelerated (but not rotating) coordinate system.

The interference pattern under these conditions will show a dependence not only 
on the mass of the respective particles, i.e., once again at the quantum level gravity is not  
a purely geometric effect, but it also depends on some parameters that appear in RPIF 
and therefore renders the possibility of confronting its theoretical predictions with 
a generalized version of COW experimental construction. 
Noninertial effects also emerge.
\bigskip

\section{Generalized Gravity--Induced Interference.}
\bigskip
 
Let us consider the case of a particle with mass $m$ located in a region where 
the Earth's gravitational field can be considered homogeneous, i.e., the gravitational 
acceleration $g$ is a constant. Then the motion equation along the vertical direction $Z$ with respect to an inertial reference frame is given by 
\bigskip

\begin{equation}
m\ddot{Z} + mg = 0.
\end{equation}
\bigskip

Now we observe this particle using a reference frame which has an acceleration 
along the vertical direction with respect to our previous inertial coordinate 
equal to $g[f(t) - 1]$, where $f(t)$ is an arbitrary function of time. Denoting by $z$ the vertical coordinate on this accelerated reference system 
we obtain as motion equation the following expression

\begin{equation}
m\ddot{z} + mgf(t) = 0,
\end{equation}
\bigskip

and the corresponding Lagrangian is [9]

\begin{equation}
L = {1\over 2}m(\dot{z})^2 - mgf(t)z + {1\over 2}m\Bigl((\dot{x})^2 + (\dot{y})^2\Bigr).
\end{equation}
\bigskip

At this point we introduce a generalized version of COW experiment [3]. 
A beam of particles is split into two parts and then brought together. 
Each one of these two split beams follows a different path along the vertical 
direction $z$. We do not assume that the wave packets are much smaller than the 
dimension of the loop formed by the alternate paths. We also monitor continuously 
the $z$ coordinate of each one of these two beams, and the whole experimental construction is at rest with respect 
to this accelerated reference frame that we have just introduced.

Under these conditions the associated restricted path integral for each one of the beams is 

\begin{equation}
U_{[c(t)]}(z_2, z_1) = \int_{\Omega}\omega_{[c(t)]}[z(\tau)]d[z(\tau)]exp(iS[z(\tau)]/\hbar),
\end{equation}
\bigskip

\begin{equation}
S = \int_{\tau '}^{\tau ''}[{1\over 2}m(\dot{z})^2 - mgf(t)z]d\tau.
\end{equation}
\bigskip

Here $z_1$ and $z_2$ are the endpoints of the motion along the vertical direction, $\omega_{[c(t)]}[z(\tau)]$ is the corresponding weight functional 
which contains all the information concerning the measuring apparatus, $c(t)$ the resulting measured trayectory, 
and $\Omega$ the set of all functions such that $z(\tau ') =z_1$ and $z(\tau '') = z_2$. Of course, each beam has its particular $c(t)$. 
The propagator contains also a contribution coming from $x$ and $y$ but it plays no role in this analysis.

At this point, in order to obtain theoretical predictions, we must choose a par\-ti\-cu\-lar expression 
for $\omega_{[c(t)]}[z(\tau)]$. We know that the results coming from a Heaveside weight functional [10] and those 
coming from a gaussian one [11] coincide up to the order of magnitude. This last result allows us 
to consider as our weight functional a gaussian expression. 
But a sounder justification of this choice comes from the fact that there 
are measuring processes in which the weight functional has precisely a gaussian form [12]. 
In consequence we could think about a measuring device whose weight functional is very close to a gaussian behavior.

Thus we have that 

\begin{equation}
\omega_{[c(t)]}[z(\tau)] = exp\{-{2\over T\Delta c^2}\int _{\tau '}^{\tau ''}[z(\tau) - c(\tau)]^2d\tau\}.
\end{equation}
\bigskip

Here $T = \tau '' - \tau'$ and $\Delta c$ represents the error in the position measuring.

The path integral of each beam is then
\bigskip

{\setlength\arraycolsep{2pt}\begin{eqnarray}
U_{[c(t)]}(z_2, z_1) = \int_{\Omega}d[z(\tau)]exp\{{i\over\hbar}\int_{\tau '}^{\tau ''}[{1\over 2}m(\dot{z})^2 - mgf(t)z]d\tau\}\times\nonumber\\
exp\{-{2\over T\Delta c^2}\int _{\tau '}^{\tau ''}[z(\tau) - c(\tau)]^2d\tau\}.
\end{eqnarray}}
\bigskip

This path integral is easily calculated [4] 
\bigskip

\begin{equation}
U_{[c(t)]}(z_2, z_1) = exp\Bigl(-2{<c^2>\over\Delta c^2}\Bigr)\sqrt{{mw\over 2\pi i\hbar sin(wT)}}exp\Bigl({i\over\hbar}S\Bigr).
\end{equation}
\bigskip

Here $exp\Bigl(-2{<c^2>\over\Delta c^2}\Bigr) = {1\over T}\int_{\tau '}^{\tau ''}c^2(t)dt$, 
and $S$ is the classical action of a driven complex harmonic oscillator [13] defined by 
${1\over 2}m(\dot{z})^2 - {m\over 2}w^2z^2 + F(t)z = 0$, where $F(t) = -mgf(t) -i{4\hbar\over T\Delta c^2}c(t)$ and $w = \sqrt{-i{4\hbar\over mT\Delta c^2}}$.
\bigskip

{\setlength\arraycolsep{2pt}\begin{eqnarray}
S = {mw\over 2sin(wT)}\{[z_1^2 + z_2^2]cos(wT) -2z_1z_2  \nonumber\\
+ {2z_2\over mw}\int_{\tau '}^{\tau ''}F(t)sin[w(t - \tau')]dt \nonumber\\
- {2z_1\over mw}\int_{\tau '}^{\tau ''}F(t)sin[w(t - \tau '')]dt \nonumber\\
- {2\over (mw)^2}\int_{\tau '}^{\tau ''}dt\int_{\tau '}^{t}dsF(t)\times \nonumber\\
sin[w(\tau '' - t)]F(s)sin[w(s - \tau ')]\}. 
\end{eqnarray}} 
\bigskip

\section{Interference Terms.}
\bigskip

The resulting interference pattern at point $z_2$, which appears when the two beams are recombined is given by

\begin{equation}
|U_{[a(t)]}(z_2, z_1) + U_{[b(t)]}(z_2, z_1)|^2.
\end{equation}

Here we denote the measurement outputs for the $z$ coordinate of the beams by $a(t)$ and $b(t)$.
If we re--write the actions of the two complex oscillators as $S_{[a(t)]} = S^{(1)}_{[a(t)]} + iS^{(2)}_{[a(t)]}$ and 
$S_{[b(t)]} = S^{(1)}_{[b(t)]} + iS^{(2)}_{[b(t)]}$, where $S^{(1)}_{[b(t)]}$, $S^{(2)}_{[b(t)]}$, 
$S^{(1)}_{[a(t)]}$, and $S^{(2)}_{[a(t)]}$ are real functions, then the interference term 
becomes

\begin{equation}
I = cos\Bigl({1\over \hbar}[S^{(1)}_{[a(t)]}- S^{(1)}_{[b(t)]}]\Bigr).
\end{equation}
\bigskip

After a lengthy calculation we find that we may regroup the terms in $S^{(1)}_{[a(t)]}- S^{(1)}_{[b(t)]}$ 
in five different contributions ($S^{(1)}_{[a(t)]}- S^{(1)}_{[b(t)]} = I_1 + I_2 + I_3 + I_4 + I_5$). The first two components are
\bigskip

{\setlength\arraycolsep{2pt}\begin{eqnarray}
I_1 = [z_1^2 + z_2^2]\sqrt{{m\over 2\hbar T}}\Bigl({1- e^{-4\theta} + 2e^{-2\theta}sin(2\theta)\over 
1+ e^{-4\theta} - 2e^{-2\theta}cos(2\theta)}{1\over \Delta a} \nonumber\\
- {1- e^{-4\rho} + 2e^{-2\rho}sin(2\rho)\over 
1+ e^{-4\rho} - 2e^{-2\rho}cos(2\rho)}{1\over \Delta b}\Bigr),
\end{eqnarray}} 
\bigskip

{\setlength\arraycolsep{2pt}\begin{eqnarray}
I_2 = -\sqrt{{8m\over \hbar T}}z_1z_2\Bigl({(1 - e^{-2\theta})cos(\theta) + (1 + e^{-2\theta})sin(\theta)\over 
e^{\theta}[1+ e^{-4\theta} - 2e^{-2\theta}cos(2\theta)]}{1\over \Delta a} \nonumber\\
- {(1 - e^{-2\rho})cos(\rho) + (1 + e^{-2\rho})sin(\rho)\over 
e^{\rho}[1+ e^{-4\rho} - 2e^{-2\rho}cos(2\rho)]}{1\over \Delta b}\Bigr).
\end{eqnarray}} 
\bigskip

Here we have that $\theta = \sqrt{{2\pi\hbar T\over m\Delta a^2}}$ and $\rho = \sqrt{{2\pi\hbar T\over m\Delta b^2}}$. 
$I_1$ and $I_2$ show clearly, once again, that at the quantum level gravity is not a purely geometric effect. 
Indeed, here we have the combination $m/\hbar$, as in the usual case [1, 3]. 
But these contributions to the interference pattern depend also on $\Delta a$ and on 
$\Delta b$. This fact means that if we measure the $z$ coordinate of the beams using two different 
devices (the corresponding measuring errors are not the same, i.e., $\Delta a \not = \Delta b$), then this 
difference will render a nonvanishing contribution to the interference pattern. This last 
result could allow us to test the theoretical predictions of RPIF against the 
measurements outputs coming from a generalized version of COW experiment.

The second pair of interference terms is
\bigskip

{\setlength\arraycolsep{2pt}\begin{eqnarray}
I_3 = -{m\over\hbar}{z_1e^{-\theta}\over 1+ e^{-4\theta} - 2e^{-2\theta}cos(2\theta)}
\{(1 - e^{-2\theta})cos(\theta)\times \nonumber\\ 
\int_{\tau '}^{\tau ''}e^{\gamma/\sqrt{2}}[{4\hbar\over mT\Delta a^2}a(t)sin(\gamma/\sqrt{2})
(1 + e^{-2\gamma/\sqrt{2}}) \nonumber\\
- gf(t)cos(\gamma/\sqrt{2})(1 - e^{-2\gamma/\sqrt{2}})]dt \nonumber\\
- (1 + e^{-2\theta})sin(\theta)\times \nonumber\\ 
\int_{\tau '}^{\tau ''}e^{\gamma/\sqrt{2}}[{4\hbar\over mT\Delta a^2}a(t)cos(\gamma/\sqrt{2})
(1 - e^{-2\gamma/\sqrt{2}}) \nonumber\\
+ gf(t)sin(\gamma/\sqrt{2})(1 + e^{-2\gamma/\sqrt{2}})]dt\} \nonumber\\
+ {m\over\hbar}{z_1e^{-\rho}\over 1+ e^{-4\rho} - 2e^{-2\rho}cos(2\rho)}
\{(1 - e^{-2\rho})cos(\rho)\times \nonumber\\ 
\int_{\tau '}^{\tau ''}e^{\Gamma/\sqrt{2}}[{4\hbar\over mT\Delta b^2}b(t)sin(\Gamma/\sqrt{2})
(1 + e^{-2\Gamma/\sqrt{2}}) \nonumber\\
- gf(t)cos(\Gamma/\sqrt{2})(1 - e^{-2\Gamma/\sqrt{2}})]dt \nonumber\\
- (1 + e^{-2\rho})sin(\rho)\times \nonumber\\ 
\int_{\tau '}^{\tau ''}e^{\Gamma/\sqrt{2}}[{4\hbar\over mT\Delta b^2}b(t)cos(\Gamma/\sqrt{2})
(1 - e^{-2\Gamma/\sqrt{2}}) \nonumber\\
+ gf(t)sin(\Gamma/\sqrt{2})(1 + e^{-2\Gamma/\sqrt{2}})]dt\},
\end{eqnarray}} 
\bigskip

{\setlength\arraycolsep{2pt}\begin{eqnarray}
I_4 = {m\over\hbar}{z_2e^{-\theta}\over 1+ e^{-4\theta} - 2e^{-2\theta}cos(2\theta)}
\{(1 - e^{-2\theta})cos(\theta)\times \nonumber\\ 
\int_{\tau '}^{\tau ''}e^{\mu/\sqrt{2}}[{4\hbar\over mT\Delta a^2}a(t)sin(\mu/\sqrt{2})
(1 + e^{-2\mu/\sqrt{2}}) \nonumber\\
- gf(t)cos(\mu/\sqrt{2})(1 - e^{-2\mu/\sqrt{2}})]dt \nonumber\\
- (1 + e^{-2\theta})sin(\theta)\times \nonumber\\ 
\int_{\tau '}^{\tau ''}e^{\mu/\sqrt{2}}[{4\hbar\over mT\Delta a^2}a(t)cos(\mu/\sqrt{2})
(1 - e^{-2\mu/\sqrt{2}}) \nonumber\\
+ gf(t)sin(\mu/\sqrt{2})(1 + e^{-2\mu/\sqrt{2}})]dt\} \nonumber\\
- {m\over\hbar}{z_2e^{-\rho}\over 1+ e^{-4\rho} - 2e^{-2\rho}cos(2\rho)}
\{(1 - e^{-2\rho})cos(\rho)\times \nonumber\\ 
\int_{\tau '}^{\tau ''}e^{\nu/\sqrt{2}}[{4\hbar\over mT\Delta b^2}b(t)sin(\nu/\sqrt{2})
(1 + e^{-2\nu/\sqrt{2}}) \nonumber\\
- gf(t)cos(\nu/\sqrt{2})(1 - e^{-2\nu/\sqrt{2}})]dt \nonumber\\
- (1 + e^{-2\rho})sin(\rho)\times \nonumber\\ 
\int_{\tau '}^{\tau ''}e^{\nu/\sqrt{2}}[{4\hbar\over mT\Delta b^2}b(t)cos(\nu/\sqrt{2})
(1 - e^{-2\nu/\sqrt{2}}) \nonumber\\
+ gf(t)sin(\nu/\sqrt{2})(1 + e^{-2\nu/\sqrt{2}})]dt\}.
\end{eqnarray}} 
\bigskip

In these last two expressions we have introduced the following definitions 
$\gamma = \sqrt{{4\hbar\over mT\Delta a^2}}(t - \tau'')$, $\Gamma = \sqrt{{4\hbar\over mT\Delta b^2}}(t - \tau'')$, 
$\mu = \sqrt{{4\hbar\over mT\Delta a^2}}(t - \tau')$, $\nu = \sqrt{{4\hbar\over mT\Delta b^2}}(t - \tau')$. 
Once again we may see that the factor $m/\hbar$ appears in scene. 
The effects of performing the experiment on a noninertial reference frame emerge in these two 
expressions, i.e., $f(t)$ is present in both of them. 
If the alternate paths are not the same, then an additional term contributes to the interference pattern, i.e., $a(t)$ and $b(t)$ appear explicitly. Of course, as 
in $I_1$ and $I_2$, we have a dependence on the measuring errors $\Delta a$ and $\Delta b$, a 
difference in the measuring errors is an interference source.
 
The presence of $f(t)$ and the fact that $a(t) \not = b(t)$ render a second possibility to confront RPIF theoretical predictions against a generalized COW experiment. 
\bigskip

The last contribution to the interference term is
\bigskip

{\setlength\arraycolsep{2pt}\begin{eqnarray}
I_5 = {m\over\hbar}{Te^{-\theta}\over\theta
[1+ e^{-4\theta} - 2e^{-2\theta}cos(2\theta)]}\{-[(1 - e^{-2\theta})cos(\theta) \nonumber\\
- 
(1 + e^{-2\theta})sin(\theta)]\times
\int_{\tau '}^{\tau ''}\int_{\tau '}^{t}{e^{(\sigma + \epsilon)/2}\over 2}
\Bigl([g^2(f(t) + f(s)) \nonumber\\
- ({4\hbar\over mT\Delta a^2})^2(a(t) + a(s))][cos({\epsilon - \sigma\over\sqrt{2}})\times \nonumber\\
(e^{-\sqrt{2}\sigma} + e^{-\sqrt{2}\epsilon}) - cos({\epsilon + \sigma\over\sqrt{2}})(1 + e^{-\sqrt{2}(\sigma + \epsilon)})] \nonumber\\
+{4\hbar\over mT\Delta a^2}g[f(t)a(s) + f(s)a(t)][sin({\epsilon + \sigma\over\sqrt{2}})(1 - e^{-\sqrt{2}(\sigma + \epsilon)}) 
\nonumber\\
+ (-e^{-\sqrt{2}\sigma} + e^{-\sqrt{2}\epsilon})
sin({\epsilon - \sigma\over\sqrt{2}})]\Bigr)dtds + [(1 - e^{-2\theta})cos(\theta) \nonumber\\
+ (1 + e^{-2\theta})sin(\theta)]
\int_{\tau '}^{\tau ''}\int_{\tau '}^{t}{e^{(\sigma + \epsilon)/2}\over 2}
\Bigl([g^2(f(t) + f(s)) \nonumber\\
- ({4\hbar\over mT\Delta a^2})^2(a(t) + a(s))][sin({\epsilon - \sigma\over\sqrt{2}})
(-e^{-\sqrt{2}\sigma} + e^{-\sqrt{2}\epsilon}) \nonumber\\
+ sin({\epsilon + \sigma\over\sqrt{2}})(1 - e^{-\sqrt{2}(\sigma + \epsilon)})]\nonumber\\
-{4\hbar\over mT\Delta a^2}g[f(t)a(s) + f(s)a(t)][-cos({\epsilon + \sigma\over\sqrt{2}})(1 + e^{-\sqrt{2}(\sigma + \epsilon)})\nonumber\\
+ (e^{-\sqrt{2}\sigma} + e^{-\sqrt{2}\epsilon})
cos({\epsilon - \sigma\over\sqrt{2}})]\Bigr)dtds\}\nonumber\\
-{m\over\hbar}{Te^{-\rho}\over\rho
[1+ e^{-4\rho} - 2e^{-2\rho}cos(2\rho)]}\{-[(1 - e^{-2\rho})cos(\rho) \nonumber\\
- 
(1 + e^{-2\rho})sin(\rho)]\times
\int_{\tau '}^{\tau ''}\int_{\tau '}^{t}{e^{(\alpha + \beta)/2}\over 2}
\Bigl([g^2(f(t) + f(s)) \nonumber\\
- ({4\hbar\over mT\Delta b^2})^2(b(t) + b(s))][cos({\beta - \alpha\over\sqrt{2}})\times \nonumber\\
(e^{-\sqrt{2}\alpha} + e^{-\sqrt{2}\beta}) - cos({\beta + \alpha\over\sqrt{2}})(1 + e^{-\sqrt{2}(\alpha + \beta)})] \nonumber\\
+{4\hbar\over mT\Delta b^2}g[f(t)b(s) + f(s)b(t)][sin({\beta + \alpha\over\sqrt{2}})(1 - e^{-\sqrt{2}(\alpha + \beta)}) 
\nonumber\\
+ (-e^{-\sqrt{2}\alpha} + e^{-\sqrt{2}\beta})
sin({\beta - \alpha\over\sqrt{2}})]\Bigr)dtds + [(1 - e^{-2\rho})cos(\rho) \nonumber\\
+ (1 + e^{-2\rho})sin(\rho)]
\int_{\tau '}^{\tau ''}\int_{\tau '}^{t}{e^{(\alpha + \beta)/2}\over 2}
\Bigl([g^2(f(t) + f(s)) \nonumber\\
- ({4\hbar\over mT\Delta b^2})^2(b(t) + b(s))][sin({\beta - \alpha\over\sqrt{2}})
(-e^{-\sqrt{2}\alpha} + e^{-\sqrt{2}\beta}) \nonumber\\
+ sin({\beta + \alpha\over\sqrt{2}})(1 - e^{-\sqrt{2}(\alpha + \beta)})]\nonumber\\
-{4\hbar\over mT\Delta b^2}g[f(t)b(s) + f(s)b(t)][-cos({\beta + \alpha\over\sqrt{2}})(1 + e^{-\sqrt{2}(\alpha + \beta)})\nonumber\\
+ (e^{-\sqrt{2}\alpha} + e^{-\sqrt{2}\beta})
cos({\beta - \alpha\over\sqrt{2}})]\Bigr)dtds\}.
\end{eqnarray}} 
\bigskip
 
In this last expression we have defined $\epsilon = \sqrt{{4\hbar\over mT\Delta a^2}}(\tau '' -t)$, 
$\sigma = \sqrt{{4\hbar\over mT\Delta a^2}}(s - \tau ')$, $\alpha = \sqrt{{4\hbar\over mT\Delta b^2}}(s - \tau ')$, 
and finally $\beta = \sqrt{{4\hbar\over mT\Delta b^2}}(\tau '' - t)$. Once again the term $m/\hbar$ emerges, as well as the dependence on 
the measuring errors $\Delta b$ and $\Delta a$ and on the alternate paths $a(t)$ and $b(t)$. 
Here a new feature appears, namely there are terms in $I_5$ which result from the multiplication 
of $f(t)$ and the alternate paths. 
\bigskip

\section{Conclusions.}
\bigskip

We have analyzed, in the context of RPIF, a possible generalization of COW gravity--induced interference 
experiment in which the vertical coordinate of the two involved split beams is continuously measured. 
Mass emerges in the interference terms once again in the form $m/\hbar$, i.e., gravity is at the quantum level, as was already shown in COW experiment, 
not a purely geometric effect. 
We have also seen that interference depends not only on the alternate paths of the 
split beams, as in the original experiment, but also on the measuring error of 
the experimental devices. In other words, if we measure the vertical coordinate of the beams with experimental 
devices that have di\-ffe\-rent measuring errors, then this difference will be an interference source. 

It must also be mentioned that using Hawking's approach no quantum gravity effect appears in connection 
with the here discussed problem [14].

Summing up, from this analysis we could conclude that RPIF theoretical predictions 
could be confronted against the measurement outputs of a generalized version of COW gravity--induced interference experiment.   

\bigskip
 
\Large{\bf Acknowledgments.}\normalsize
\bigskip

The author would like to thank A. Camacho--Galv\'an and A. A. Cuevas--Sosa for their 
help, and D.-E. Liebscher for the fruitful discussions on the subject. 
The hospitality of the Astrophysikalisches Institut Potsdam is also kindly acknowledged. 
This work was supported by CONACYT Posdoctoral Grant No. 983023.

\end{document}